\documentclass[prl,twocolumn,showpacs,preprintnumbers]{revtex4-1}
\usepackage[version=3]{mhchem} 

\usepackage{graphicx} 
\usepackage{dcolumn}
\usepackage{bm}
\usepackage{booktabs} 
\usepackage{amsmath}  
\usepackage[colorlinks=true,linkcolor=red]{hyperref} 
\usepackage{memhfixc} 

\begin{document}

\title{Determination of substrate pinning in epitaxial and supported graphene layers via Raman scattering}
\author{Nicola Ferralis}
\email{feranick@gmail.com}
\affiliation{Department of Chemical and Biomolecular Engineering, University of California,
 Berkeley, California 94720}
\altaffiliation{Current address: Department of Materials Science and Engineering, MIT, MA 02139}  
\author{Roya Maboudian}
\affiliation{Department of Chemical and Biomolecular Engineering, University of California,
 Berkeley, California 94720}
\author{Carlo Carraro}
\affiliation{Department of Chemical and Biomolecular Engineering, University of California,
 Berkeley, California 94720}

\date{\today}

\begin{abstract}
The temperature-induced shift of the Raman G line in epitaxial graphene on SiC and Ni surfaces, as well as in graphene supported on SiO$_2$, is investigated with Raman spectroscopy. The thermal shift rate of epitaxial graphene on 6H-SiC(0001) is found to be about three times that of freestanding graphene. This result is explained quantitatively as a consequence of pinning by the substrate. In contrast, graphene grown on polycrystalline Ni films is shown to be unpinned, i.e., to behave elastically as freestanding, despite the relatively strong interaction with the metal substrate. Moreover, it is shown that the transfer of exfoliated graphene layers onto a supporting substrate can result in pinned or unpinned layers, depending on the transfer protocol.

\end{abstract}


\maketitle

Graphene films on insulating substrates have great potential for realization of high speed electronic devices \cite{Schwierz_NNano2010}, as demonstrated recently by graphene transistor operation near terahertz frequencies \cite{Lin_S2010}. Compared to research into graphene's remarkable electronic properties, much less is known about its mechanical properties. Yet, the response of graphene films to mechanical stimuli is an important fundamental topic, with potentially far reaching applications. Noteworthy among these is strain engineering, which offers the tantalizing prospect of manipulating graphene's electromagnetic properties \cite{Guinea_NP2009, Levy_S2010}. Moreover, mechanical energy transfer processes dominate the response (and hence, dictate the design) of graphene-based nanomechanical transducers and actuators~\cite{Bunch_S315}. Lastly, the success of  many promising graphene-on-insulator fabrication processes hinges on suitable mechanical manipulation of single layers of material~\cite{Kim_N2009,Reina_NL2009,Ruoff_S2009}. 

Modeling of a graphene film as an elastic continuum can offer much valuable insight into its mechanical response \cite{Kim_EPL2008, Los_PRB2009}. The constitutive ingredients of such model, namely Young's modulus and bending rigidity, are known experimentally~\cite{Frank_JVSTB2007}. However boundary conditions must also be specified in order to solve the model. These are determined by the complex interactions between graphene film and substrate, which ultimately result into pinned or unpinned films. A pinned layer implies continuity of tangential displacements, while an unpinned layer implies zero tangential stress.  

In this Letter, we show that information about film pinning is obtained unambiguously from measurements of the temperature dependence of the Raman G line position, which allow one to separate out ``spurious" effects, like charge or strain, that dominate the absolute line position at any given temperature. The significance of the result is illustrated in three different systems. Growth by sublimation on SiC(0001) is shown to yield pinned films. Growth by chemical vapor deposition on Ni films yields unpinned films. Finally, transfer of an exfoliated film onto an oxidized Si substrate results in either pinned or unpinned films depending on the transfer protocol. These results may seem surprising at first, given that  graphene on SiC is known to grow on top of a ``buffer'' layer, which decouples it from the substrate \cite{Chen_SS2005, Mattausch_PRL2007, Varchon_PRL2007, Emtsev_PRB2008}, while on metals, strong interactions involving hybridization of electronic orbitals are believed to exist \cite{Wu_JAP2010}. The explanation of this apparent paradox resides likely in the nature of the potential energy landscape on semiconductor vs metal surfaces, the former having a highly corrugated potential, the latter a much smoother one. It is the strength of the lateral, rather than the vertical, interactions between film and substrate that is responsible for pinning. 
\begin{figure}
\includegraphics[width=8.4cm]{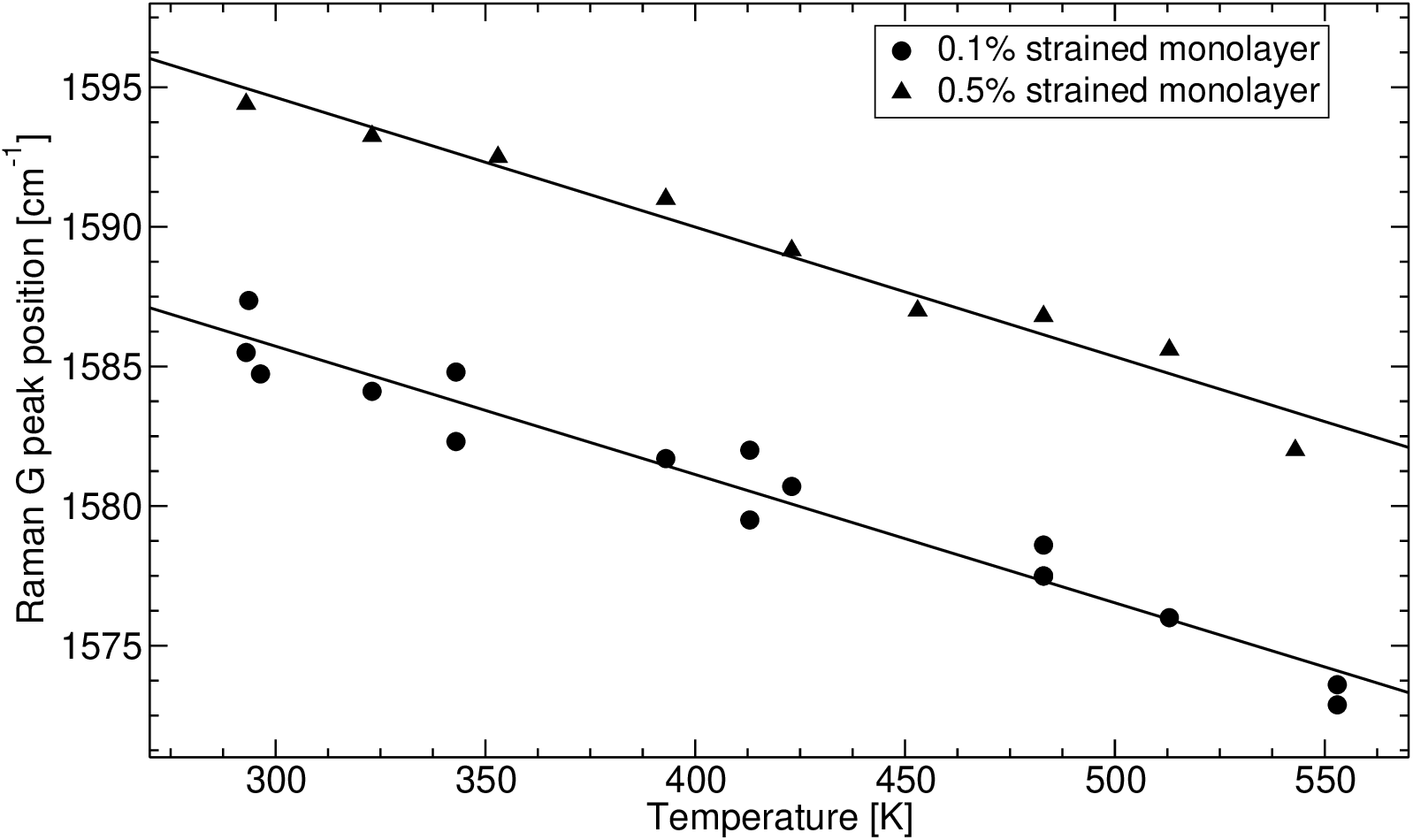}
\caption{\label{Fig1} Raman thermal line shift of the G vibrational mode for epitaxial graphene films, prepared with different amounts of strain on 6H-SiC(0001).}
\end{figure}

Figure~\ref{Fig1} shows the values of the frequency of the zone-center optical phonon of graphene (the G line) \cite{Ferrari_SSC2007,Malard_PR2009} recorded on two epigraphene monolayer samples grown in ultrahigh vacuum by sublimation of Si from the (0001) face of a 6H-SiC crystal under Si flux~\cite{foot1}.  Note that epigraphene films on SiC can be deliberately produced with different amounts of internal stress at room temperature, and hence, the G line frequency at room temperature could take on any value between 1580 and 1605 cm${}^{-1}$ in a given sample, depending on sample preparation \cite{Ferralis_PRL2008}. The samples in Fig.~\ref{Fig1}  had strain values of 0.1\% and 0.5\% at room temperature. Previous measurements on freestanding graphene monolayers produced by exfoliation from HOPG crystals \cite{Calizo_NL2007,Calizo_APL2007} observed an average redshift rate of -0.016 cm${}^{-1}$/K in the temperature range 100~K~$<T<$~300~K, in agreement with theoretical calculations on 2D graphene \cite{Bonini_PRL2007}. The redshift rates observed in epigraphene are almost three times as large, regardless of the amount of internal stress in the films observed at room temperature. Thus, the $slope$ of the curves is independent of the initial amount of stress present in the film, while the absolute position of the G line can vary for several reasons (such as engineered strain \cite{Huang_PNAS2009, Mohiuddin_PRB2009} or doping \cite{Pisana_NM2007, Das_NN2008}), which are decoupled from the thermal redshift. This implies that a single measurement at constant temperature is not sufficient to determine whether a film is pinned to the substrate or not. Pinning is instead ascertained by performing measurements at different temperatures.

To understand the behavior of the thermal line shift, recall that at constant pressure, the rate of change of phonon frequency  with temperature is given by the sum of two terms, 
\begin{equation}
\left(\frac{d\omega}{dT}  \right)_P=\chi_T(T)+\chi_a(T)
\label{eq1}
\end{equation}
where:
\begin{equation}
\chi_T(T)=\left( \frac{\partial \omega}{\partial a} \right)_T~\left(\frac{\partial a}{\partial T}\right)_P;
~~~~~\chi_a(T)=\left(\frac{\partial \omega}{\partial T}\right)_a.
\label{eq2}
\end{equation}

Phonon frequencies depend on temperature in two ways, explicitly and implicitly through the lattice constant $a$. The explicit dependence is contained in the constant specific area term $\chi_a$, which is due to many-phonon interactions: it reflects the anharmonicity of the C-C potential in the graphene lattice. The implicit dependence, $\chi_T$, is given by the product of the rate of change of frequency with lattice constant $a$, times the rate of lattice thermal expansion. It is this latter term that is affected by pinning of the layer to the substrate, i.e., by boundary conditions at the substrate-film interface. For the pinned layer, we assume continuity of tangential displacements, and calculate the frequency shift of the G line in epigraphene by using the coefficient of linear thermal expansion (CTE) of SiC, $\alpha_{\rm SiC}$ \cite{Slack_JAP1975}, but keeping for all other terms the {\it ab initio} values appropriate for freestanding graphene (denoted by the the subscript ``free'' and taken from ref.~\cite{Bonini_PRL2007}):
\begin{equation}
\begin{split}
\omega(T)=\omega(T_0)+\int_{T_0}^{T}dT' \chi_{a, {\rm free}}(T')\\
+\int_{T_0}^{T}dT' \chi_{T, {\rm free}}(T')\frac{\alpha_{\rm SiC}(T')}{\alpha_{\rm free}(T')}.
\end{split}
\label{eq3}
\end{equation}

\begin{figure}
\includegraphics[width=8.6cm]{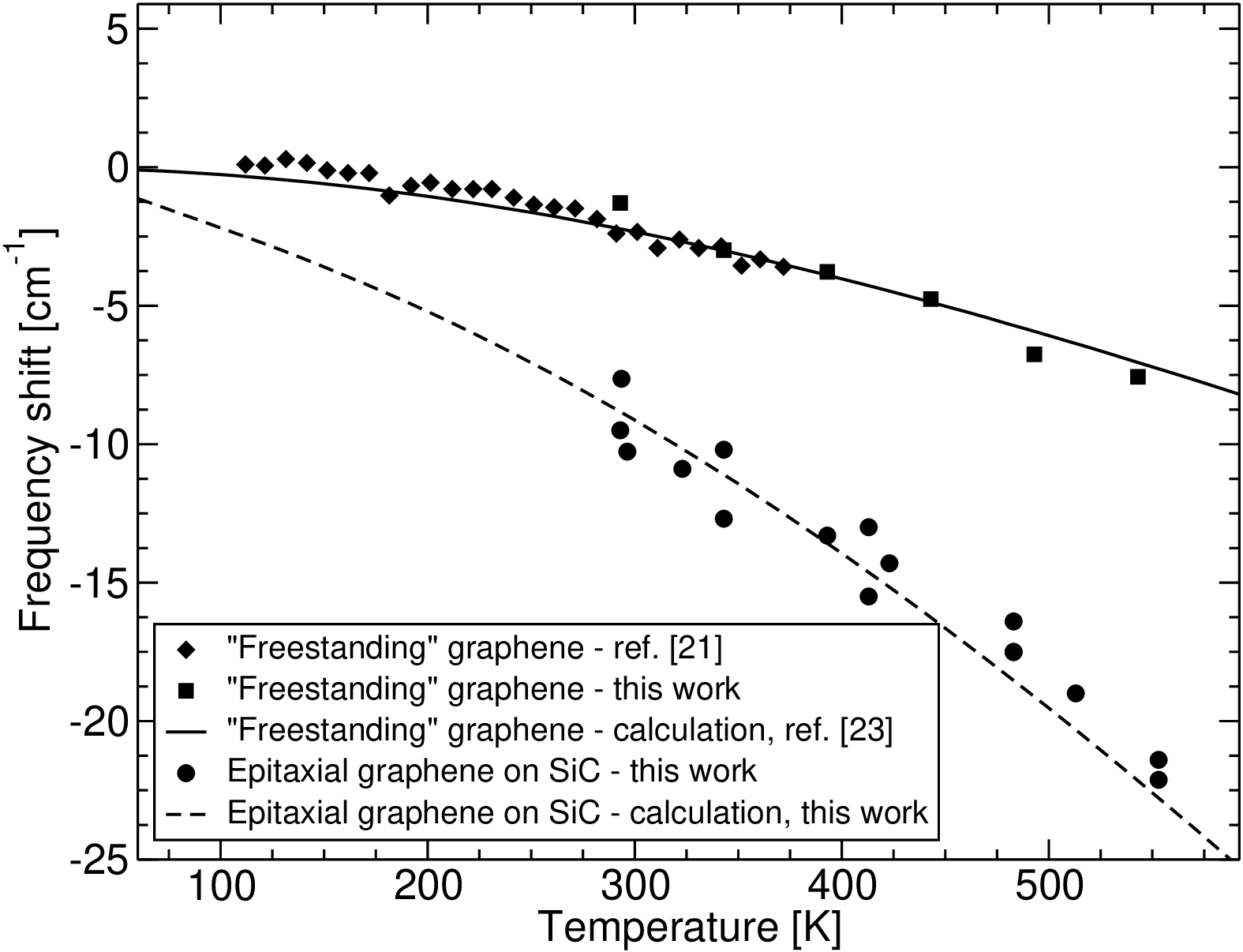}
\caption{\label{Fig2} Comparison between theoretical and experimental thermal G-line shifts for freestanding and epitaxial graphene films (re-plotted from Fig.~\ref{Fig1}). The thermal line shifts in both the theoretical curves and the experimental data are referred to the position of the G line at absolute zero.}
\end{figure}

Note that the integration constant, $\omega(T_0)$ (where $T_0$ is an arbitrary reference temperature), need not coincide with the line position of a freestanding film in equilibrium: its value may be shifted due to engineered stress or charge effects, for example.
The results of the calculations are shown in Fig.~\ref{Fig2}, along with the data. For comparison, we also plot the calculated thermal G-line shift of free graphene \cite{Bonini_PRL2007}, along with experimental measurement from ref.~\cite{Calizo_NL2007}, which we have augmented here by performing measurements on free films at higher temperatures.

The agreement between theory and experiment confirms that on the SiC(0001) surface, substrate effects are entirely accounted for by using the boundary conditions appropriate for a pinned monolayer. However, as shown below, this is not to be regarded as a forgone conclusion: we have observed strikingly different behavior on different substrates. 

\begin{figure}
\includegraphics[width=8.6cm]{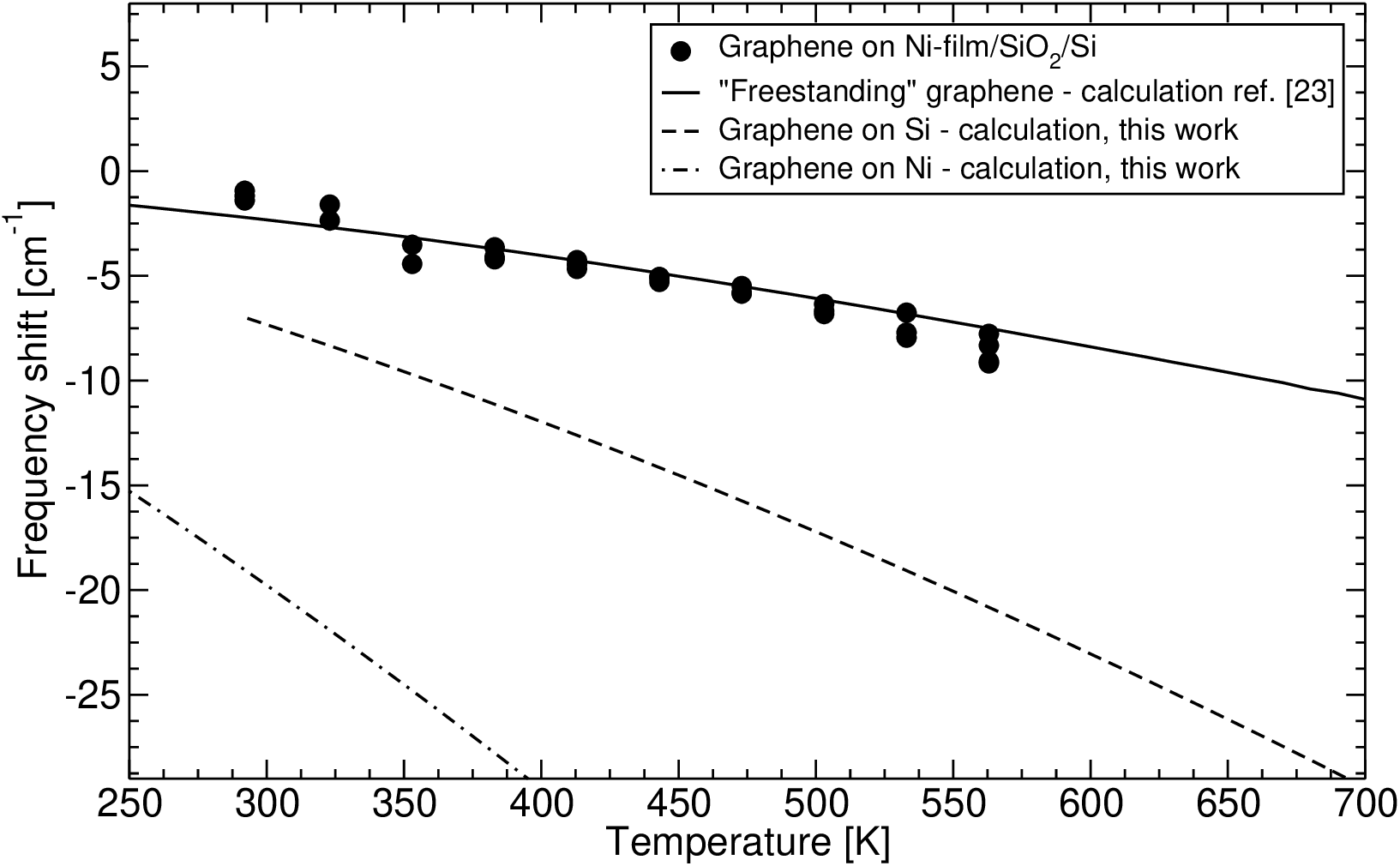}
\caption{\label{Fig3} The experimental thermal line shifts of the Raman G line for a monolayer graphene grown via CVD on a Ni/SiO$_2$/Si substrate is compared to the calculated values of the thermal shifts for freestanding graphene, graphene pinned to Si, and graphene pinned to Ni (from  Eq.~\ref{eq3}, using the linear thermal expansion coefficient of Si~\cite{Watanabe_IJTP2004} and Ni~\cite{Nix_PR1941}, respectively.)}
\end{figure}

Graphene monolayer growth by chemical vapor deposition (CVD) has been demonstrated on several metal surfaces, including Ni substrates \cite{Kim_N2009}, where strong film-substrate interactions have been observed \cite{Dedkov_PRL2008, Gruneis_PRB2008}. In some cases, interactions lead to the suppression of the Raman signal, possibly owing to the hybridization between the metal \textit{d} bands and the carbon \textit{p$_z$} bands \cite{Allard_NL2010}. On polycrystalline Ni films, perfect epitaxy is not expected, and in fact, a Raman signal is always observed \cite{Kim_N2009}. We have measured the thermal line shifts for graphene monolayers grown by CVD on thin Ni films (300~nm) on SiO$_2$/Si substrates~\cite{foot2}. The results are shown in Fig.~\ref{Fig3}, along with calculated values for freestanding graphene \cite{Bonini_PRL2007}, for graphene on Si, and for graphene on bulk Ni. We emphasize that the mechanical behavior of a well adhered thin film (or thin film stack) on a thick substrate is dictated by the properties of the bulk substrate. When comparing to our experiment, the relevant CTE to be used in Eq.~\ref{eq3} is that of Si~\cite{Watanabe_IJTP2004}, since the Si wafer thickness is 500~$\mu$m vs. 300 nm Ni film thickness. The good agreement between the experimental data and the theoretical curve for freestanding graphene, rather than for graphene on Si substrate, indicates that on polycrystalline Ni films, graphene is completely unpinned from the substrate and it behaves essentially like a freestanding layer.  This free-like behavior can be rationalized by considerations of surface potential corrugation, which is usually much smaller on metal surfaces compared to insulators (or semiconductors). Since film pinning involves tangential displacements, it is plausible that graphene monolayers grown on metal surfaces, are free to slide laterally, the rather strong electronic interactions with the substrate notwithstanding.   

\begin{figure}
\includegraphics[width=8.6cm]{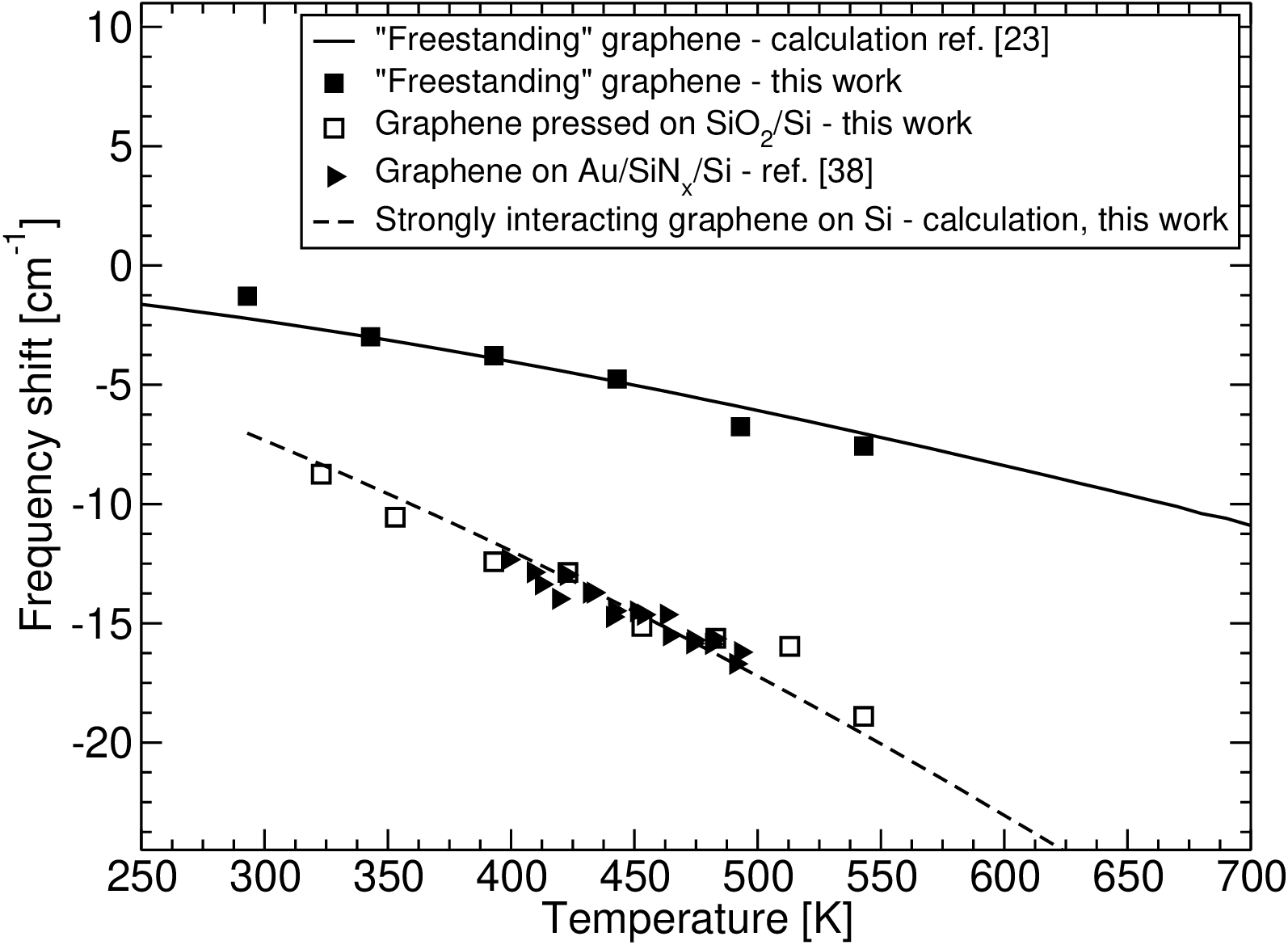}
\caption{\label{Fig4} Experimental thermal line shifts of the Raman G-line for freestanding graphene and for graphene films pressed onto a SiO$_2$ and Au/SiN$_x$~\cite{Cai_NL2010} substrates. }
\end{figure}

Other fabrication methods involve transfer of graphene layers onto an inert substrate, usually an insulator~\cite{Kim_N2009,Reina_NL2009,Ruoff_S2009}. The question then arises whether these supported layers are free or pinned. Interestingly, two different transfer methods lead consistently to contrasting answers.  
In the first method, we produced graphene films by mechanical exfoliation of HOPG crystals \cite{Novoselov_S2004}; the substrate die (a 300~nm thick thermal oxide film on Si) was dipped in a toluene solution where graphene flakes were dispersed. Graphene layers thus collected are referred to as ``freestanding." In the second method, graphite platelets were exfoliated from HOPG samples using Scotch$^{TM}$ tape, and immediately pressed against the SiO$_2$/Si substrate die. After sonication of the die to remove the larger flakes, $\sim$1~$\mu$m single and double layers of graphene (characterized by Raman spectroscopy~\cite{Blake_APL2007} and optical microscopy~\cite{Ni_NL2007}) remained attached to the surface. Figure~\ref{Fig4} shows the thermal line shifts measured for these two samples (solid and empty squares, respectively). For comparison, data corresponding to graphene layers pressed onto Au/SiN$_x$/Si substrates are plotted from ref.~\cite{Cai_NL2010}. A visual inspection shows very similar thermal line shifts for graphene pressed on SiO$_2$/Si and Au/SiN$_x$/Si substrates (Figure~\ref{Fig4}). The experimental data is compared with calculations for freestanding graphene and graphene pinned to a Si substrate, (solid and dashed lines in Fig.~\ref{Fig4}, respectively). Agreement between theory and experiments points out that only the films dispersed from solution behave as freestanding, whereas pressing or stamping results in pinned films. Consistent with our discussion of tangential vs vertical interactions, we believe pinning of pressed films is brought about by their better conformation to nanoscale surface roughness, as well as by the squeezing out of liquid-like layers at the film-substrate interface (e.g., physisorbed water~\cite{Chen_NL2009}). The values of $d\omega/dT$ and their corresponding temperature ranges are reported in Table~\ref{tab1}.

\begin{table}
\begin{tabular}{|c|c|c|c|c|c|}
	\hline
    Sample & $d\omega/dT$ & T range  &  Ref.  & Theory  & Ref.  \\
    & [cm$^{-1}$/K] & [K] &  &  &  \\
	 \hline
   Freestanding & -0.009$\pm$0.002 & 150-250 & \cite{Calizo_NL2007} & -0.011 & \cite{Bonini_PRL2007} \\ 
       & -0.015$\pm$0.003 & 300-400 & \cite{Calizo_NL2007}  & -0.017 & \cite{Bonini_PRL2007} \\ 
   	\hline
   Pressed on SiO$_2$/Si& -0.052$\pm$0.004 & 300-400 & $^{**}$ &-0.046$^*$ & $^{**}$ \\
   \hline
   On Au/SiN$_x$/Si & -0.040$\pm$0.002 & 400-500 & \cite{Cai_NL2010}  &-0.052$^*$ & $^{**}$ \\
   \hline
   Epi-G on SiC & -0.043$\pm$0.013 & 300-400 & $^{**}$ &-0.048 & $^{**}$ \\
   
\hline
\end{tabular}

$^*$ Calculation treats the substrate as pure silicon.

$^{**}$ This work.

\caption{Comparison between values of the thermal line-shift rates $d\omega/dT$ measured over different temperature ranges for graphene films prepared by different methods and calculated values from Eq.~\ref{eq3}. }
\label{tab1}
\end{table}

In summary, we have shown that the position of the Raman G line of graphene monolayers at a given temperature can be influenced by the graphene interaction with the substrate, resulting in thermal line shifts that can be calculated based on simple thermodynamic arguments, the only inputs being thermomechanical properties of free graphene (known, e.g., from density functional calculations) and the thermal expansion coefficient of the substrate. 
Conversely, experimental determination of the temperature shift of the G peak in the Raman spectrum allows one to determine whether a graphene film, grown or transferred onto a substrate, is pinned to it, or elastically decoupled from it. These results shed light on the diverse effects of substrate interactions in some of the most common graphene production methods.  Furthermore, they imply that pinning effects must be considered properly when using Raman spectroscopy for thermal measurements, such as in the determination of thermal conductivity \cite{Balandin_NL2008,Faugeras_ACSN2010,Cai_NL2010,Ghosh_NM2010}.

\begin{acknowledgments}
Support from the National Science Foundation (Grants EEC-0832819 and CMMI-0825531), and from DARPA-MTO is gratefully acknowledged. 
\end{acknowledgments}

\bibliographystyle{apsrev4-1}

\begin{thebibliography}{10}
\bibitem{foot2} Graphene growth on Ni substrates is carried out in a LP-CVD reactor for 5 minutes at 1000$^{\circ}$C, in a flow of 50 sccm CH$_4$, 200 sccm Ar and 65 sccm H$_2$.
\end{thebibliography}
%
\end{document}